\documentclass{article}
\usepackage{amsfonts}
\usepackage{amsmath}

\begin{document}

\begin{center}
{\large \textbf{\\[2mm]
Reductions of Topologically Massive Gravity I: }}

{\large \textbf{Hamiltonian Analysis of The Second Order Degenerate
Lagrangians}}

\bigskip

\bigskip

Filiz \c{C}a\u{g}atay-U\c{c}gun$^{1,2}$, O\u{g}ul Esen$^{3}$ and Hasan G\"{u}%
mral$^{4}$

\bigskip
\end{center}

$^{1}$Graduate School of Science and Engineering, Yeditepe University, 34755
Ata\c{s}ehir, \.{I}stanbul, Turkey

$^{2}$Department of Mathematics, I\c{s}\i k University,

$^{3}$Department of Mathematics, Gebze Technical University, 41400 Gebze,
Kocaeli, Turkey, oesen@gtu.edu.tr

$^{4}$Department of Mathematics, Australian College of Kuwait, 13015 Safat,
Kuwait, h.gumral@ack.edu.kw

\bigskip

\textbf{Abstract:} We study the Hamiltonian formalisms of the second order
degenerate Cl\`{e}ment and Sar\i o\u{g}lu-Tekin Lagrangians. The
Dirac-Bergmann constraint algorithm is employed while arriving at the total
Hamiltonian functions and the Hamilton's equations on the associated
momemtum phase spaces whereas the Gotay-Nester-Hinds algorithm is run while
investigating the Skinner-Rusk unified formalism on the proper Whitney
bundles.


\textbf{AMS2010:} 70H45, 70H50, 70H05, 83E05.

\section{Introduction}

The action for topologically massive gravity consists of the action for
cosmological gravity and the Chern-Simons term. Cl\'{e}ment, in his search
for particle like solutions for this theory, reduced the action \cite{cle92,
cle94a,cle94b} to the second order degenerate Lagrangian density
\begin{equation}
L^{C}=-\frac{m}{2}\zeta \dot{X}^{2}-\frac{2m\Lambda }{\zeta }+\frac{\zeta
^{2}}{2\mu m}\mathbf{X}\cdot (\mathbf{\dot{X}}\times \mathbf{\ddot{X}})
\label{clemlag}
\end{equation}%
depending on positions $\mathbf{X}$, velocities $\mathbf{\dot{X}}$ and
accelerations $\mathbf{\ddot{X}.}$ Here, the inner product is defined by the
Lorentzian metric, $\zeta =\zeta (t)$ is a function which allows arbitrary
reparametrization of the variable $t$ whereas $\Lambda $ and $1/2m$ are the
cosmological and Einstein gravitational constants, respectively.

In a more recent work \cite{st06}, Sar\i o\u{g}lu and Tekin considered an
action consisting of Einstein-Hilbert, Chern-Simons and Pauli-Fierz terms
and, obtained the reduced Lagrangian density%
\begin{equation}
L^{ST}=\frac{1}{2}\left[ a(\dot{X}^{2}+\dot{Y}^{2})+\frac{2}{\mu }\dot{%
\mathbf{Y}}\cdot \ddot{\mathbf{X}}-m^{2}(\mathbf{Y}^{2}+\mathbf{X}^{2})%
\right]  \label{stlag}
\end{equation}%
by suppressing the spatial part of the theory. Here, $a,\mu ,m$ are
parameters and $\mathbf{X},\mathbf{Y}$ are the position three-vectors. In
the context of higher derivative theories, they also considered
Pais-Uhlenbeck oscillator as a nonrelativistic limit. This is described by
the nondegenerate Lagrangian density%
\begin{equation}
L^{PU}[{X}]=\frac{1}{2}\left[ \ddot{X}^{2}-(q^{2}+p^{2})\dot{X}^{2}{%
+p^{2}\Omega ^{2}}X^{2}\right]  \label{pulag}
\end{equation}%
where ${X}$ is a real dynamical variable, $p$ and $q$ are positive real
parameters \cite{pu1950}. For the theory of topological massive gravity, and Hamiltonian analysis in the ADM framework we refer to the poineering works of Deser, Jackiw and Templeton \cite{DeJaTe82,DeJaTe82b}.

Our interest in the Sar\i o\u{g}lu-Tekin and Cl\`{e}ment Lagrangians will be
in the framework of the geometry of dynamical systems generated by second
order degenerate Lagrangians. Although, there exists extensive studies \cite%
{BoKo05,MaDa05,Mas16,Mas16b,Mo10} on the several aspects of the Hamiltonian
formulations of the Pais-Uhlenberg Lagrangian (\ref{pulag}), the Hamiltonian
formulations of the Sar\i o\u{g}lu-Tekin and Cl\`{e}ment Lagrangians are
absent in the literature. Sar\i o\u{g}lu-Tekin and Cl\`{e}ment Lagrangians
are degenerate in the sense of Ostrogradsky. For the degenerate or/and
constraint systems, the Legendre transformation is not possible in a
straight forward way. To achieve this, one may need to employ the
Dirac-Bergmann algorithm \cite{Be56,Dirac,Dirac1950,SuMu74} or,
equivalently, its geometric version Gotay-Nester-Hinds algorithm \cite%
{GoNe79,GoNe80,GoNe84,GoNeHi}. Here is an incomplete list \cite%
{BaGoPoRo88,CrCa86,GoRa94,GrPoRo91,KhRa96,Ku94,NaHa96,Ne89,Po89} for the
Legendre transformations of singular or/and constraint higher order
Lagrangian systems. We, additionally, refer some recent studies; \cite%
{CoPr16} for the Legendre transformation of higher order Lagrangian systems
in terms of Tulczyjew's approach, \cite{ChFaLiTo13} for the stability
problem, \cite{LeMaMaRo02} for the theory on the jet bundles, and \cite%
{CrMoRo13, CrGoMoRo16} for the detail analysis on the second order
Lagrangians whose dependence on the accelerations are linearly and/or affine.

At the beginning of 80s, Skinner and Rusk proposed a unification of
Lagrangian and Hamiltonian formalisms on the Withney product of velocity and
momentum phase spaces \cite{Sk83,SkRu83,SkRu83b}. Adaptation of the
Skinner-Rusk unified formalism for the higher order Lagrangian systems is
achieved recently by Prieto Mart\'{\i}nez and Rom\'{a}n-Roy \cite{PrRo11}.
In the literature, some other versions of the Skinner-Rusk formalism are
also available, for example, a field theoretical version is presented in
\cite{CaLeDiVa09,Vi09}, for Lie groups we refer \cite{CoDi11}, and for an
application to the control theory, see \cite{BaEcdeMuRo07}.

There are two main goals of the present paper. The first one is to obtain
the total Hamiltonian functions, the Hamilton's equations, the Dirac-Poisson
brackets for the Cl\`{e}ment Lagrangian (\ref{clemlag}) and the Sar\i o\u{g}%
lu-Tekin Lagrangian (\ref{stlag}). The second goal is to present the
Skinner-Rusk unified formalisms of these theories.

To achieve these goals, the paper is organized into three main sections. For
the sake of completeness, and in order to widen the spectrum of the
potential readers, we shall reserve the following section for some necessary
theoretical background. Accordingly, we shall start to the next section by
recalling the Ostrogradsky-Legendre transformation. It will be shown that
reparametrization invariant second order Lagrangians must be degenerate and
must have zero energy. The Dirac-Bergmann constraint algorithm and
construction of the Dirac bracket will be summarized. The following section
will be ended with a discussion on the Skinner-Rusk unified formalism.

The last two sections, namely $3$ and $4$, will be devoted for the
investigations on the Sar\i o\u{g}lu-Tekin and Cl\`{e}ment Lagrangians,
respectively. For these sections, the itinerary maps that we shall follow
are almost the same. At first, we shall identify the configuration spaces,
tangent and cotangent bundles. Then, the associated energy functions will be
written. After introducing the primary sets of constraints, the total
Hamiltonian function will be written and the Dirac-Bergmann algorithm will
be run in order to identify the final constraint submanifold. In each step
of the algorithm, we shall revise the total Hamiltonian by adding the
secondary constraints. Once the final constraint set is determined, it is
immediate to write the Hamilton's equations. This is the first and most
common way. An alternative way arriving at the Hamilton's equations is to
construct the Dirac bracket. To do this, we shall first classify the
constraints, determining the final constraint submanifold, into two classes,
namely the first and the second. Then, using this classification, we shall
define the Dirac brackets associated with the physical systems. Finally, we
shall exhibit the Skinner-Rusk unified formalisms of the Cl\`{e}ment and
Sar\i o\u{g}lu-Tekin Lagrangians. To do this, the Gotay-Nester-Hinds
algorithm will be employed on the associated Whitney bundles.

\section{Hamiltonian analysis of the second order Lagrangians}

Let $M$ be an $m$-dimensional configuration manifold $M$ with local
coordinates $\mathbf{X}=\left( X^{1},...,X^{m}\right) $. The velocity phase
space of the system is $2m$-dimesional manifold and it is the tangent bundle
$TM$ of $M$ with the induced coordinates $(\mathbf{X,\dot{X}}),$ \cite%
{AbMa78}. The second order tangent bundle $T^{2}M$ additionally includes the
accelerations $\mathbf{\ddot{X}}$ hence it can be equipped with a coordinate
system $(\mathbf{X,\dot{X},\ddot{X}})$, \cite{GrPoRo91,LeRo11,Su14}. We fix
the notation $[\mathbf{X}]$ in order to represent three vectors $(\mathbf{X,%
\dot{X},\ddot{X}})$. The third order tangent bundle $T^{3}M$ carries the
local coordinates $\mathbf{(X,\dot{X},\ddot{X},\dddot{X}).}$ If $TTM$ is
equipped with the coordinates $(\mathbf{X,V,\dot{X},\dot{V}})$, then we can
arrive at the second iterated bundle $T^{2}M$ through the identification $%
\mathbf{V=\dot{X}}$.

\subsection{Jacobi-Ostrogradsky momenta}

The history of the theory of Hamiltonian fomulations of the higher order
Lagrangian systems dated back to more than 150 years ago to the pioneering
work of Ostrogradsky \cite{ost50}.

A second order Lagrangian density $L[\mathbf{X}]=L(\mathbf{X,\dot{X},\ddot{X}%
})$ is a function on the second order tangent bundle $T^{2}M$. The
functional differential
\begin{equation}
d(L[\mathbf{X}]dt)=({\frac{\partial L}{\partial \mathbf{X}}}\cdot d\mathbf{X}%
+{\frac{\partial L}{\partial \mathbf{\dot{X}}}}\cdot d\mathbf{\dot{X}}+{%
\frac{\partial L}{\partial \mathbf{\ddot{X}}}}\cdot d\mathbf{\ddot{X}})=%
\mathcal{E}_{\mathbf{X}}(L[\mathbf{X}])\cdot d\mathbf{X}+{\frac{d}{dt}}%
\theta _{L}[\mathbf{X}]  \label{variation}
\end{equation}%
of $L[\mathbf{X}]$ consists of two terms. The first one is the
Euler-Lagrange equations given by%
\begin{equation}
\mathcal{E}_{\mathbf{X}}(L[\mathbf{X}])\equiv {\frac{\partial L}{\partial
\mathbf{X}}}-{\frac{d}{dt}}{\frac{\partial L}{\partial \mathbf{\dot{X}}}}+{%
\frac{d^{2}}{dt^{2}}}{\frac{\partial L}{\partial \mathbf{\ddot{X}}}}=0,
\label{ele}
\end{equation}%
and the second term is a boundary term which is the total derivative of the
Lagrangian one-form
\begin{equation}
\theta _{L}[\mathbf{X}]\equiv ({\frac{\partial L}{\partial \mathbf{\dot{X}}}}%
-{\frac{d}{dt}}{\frac{\partial L}{\partial \mathbf{\ddot{X}}}})\cdot d%
\mathbf{X}+{\frac{\partial L}{\partial \mathbf{\ddot{X}}}}\cdot d\mathbf{%
\dot{X}.}  \label{theta}
\end{equation}%
For Lagrangians resulting in the same Euler-Lagrange equations (\ref{ele}), $%
\theta _{L}$ is not unique. However, its functional exterior derivative
\begin{equation}
\Omega _{L}[\mathbf{X}]\equiv d\theta _{L}[\mathbf{X}]\text{ }  \label{omega}
\end{equation}%
is a well-defined presymplectic two-form on $T^{2}M$.

On the dual picture, the momentum phase space $T^{\ast }TM$ is a canonical
symplectic manifold with coordinates $(\mathbf{X,\dot{X},P}^{0},\mathbf{P}%
^{1})$ hence it is endowed with the canonical Poisson bracket which results
in the fundamental Poisson bracket relations
\begin{equation*}
\{X^{i},P_{j}^{0}\}=\{\dot{X}^{i},P_{j}^{1}\}=\delta _{j}^{i}
\end{equation*}%
and, all the others are zero. The form of the Lagrangian one-form $\theta
_{L}$ in (\ref{theta}) suggests that we can introduce the momenta
\begin{equation}
\mathbf{P}^{0}[\mathbf{X}]=\frac{\partial L}{\partial \mathbf{\dot{X}}}-%
\frac{d}{dt}\frac{\partial L}{\partial \mathbf{\ddot{X}}}\;,\;\;\;\mathbf{P}%
^{1}[\mathbf{X}]=\frac{\partial L}{\partial \mathbf{\ddot{X}}},  \label{jo}
\end{equation}%
for a second order Lagrangian as which called as the Jacobi-Ostrogradsky
momenta. In this definition, the Lagrangian one-form $\theta _{L}$ turns out
to be
\begin{equation*}
\theta _{L}[\mathbf{X}]\equiv \mathbf{P}^{0}[\mathbf{X}]\cdot d\mathbf{X}+%
\mathbf{P}^{1}[\mathbf{X}]\cdot d\mathbf{\dot{X}.}
\end{equation*}%
Note that, $\theta _{L}[\mathbf{X}]$ is the pull back of the canonical
(Liouville) one-form
\begin{equation*}
\theta _{T^{\ast }TM}=\mathbf{P}^{0}\cdot d\mathbf{X}+\mathbf{P}^{1}\cdot d%
\mathbf{\dot{X}}
\end{equation*}%
on the cotangent bundle $T^{\ast }TM$ by the Legendre map,
\begin{equation}
\mathcal{F}L:T^{3}M\longrightarrow T^{\ast }TM:\left( \mathbf{X},\mathbf{%
\dot{X}},\mathbf{\ddot{X},\dddot{X}}\right) \longrightarrow \left( \mathbf{X}%
,\mathbf{\dot{X}},\mathbf{P}^{0}\mathbf{,P}^{1}\right) .  \label{FL}
\end{equation}

\subsection{Reparametrization invariant Lagrangians}

In this subsection, we discuss the functional obstructions and energy of the
reparametrization invariant second order Lagrangians. The definitions of
momenta in Eq.(\ref{jo}) may be inspired is the energy conservation for a
second order Lagrangian. A conservation law associated with a second order
Lagrangian may be obtained by first solving the Euler-Lagrange equations for
the partial derivatives $\partial L/\partial \mathbf{X}$ and then using them
in the expression for the total derivative $dL/dt$. The resulting
conservation law
\begin{equation}
{\frac{d}{dt}}E_{L}[\mathbf{X}]=0,\;\;\;E_{L}[\mathbf{X}]\equiv \mathbf{\dot{%
X}}\cdot ({\frac{\partial L}{\partial \mathbf{\dot{X}}}}-{\frac{d}{dt}}{%
\frac{\partial L}{\partial \mathbf{\ddot{X}}}})+\mathbf{\ddot{X}}\cdot {%
\frac{\partial L}{\partial \mathbf{\ddot{X}}}}-L[\mathbf{X}]  \label{enrgy}
\end{equation}%
is a generalization of the well-known definition of the canonical energy for
first order Lagrangians. Using the definitions in Eq.(\ref{jo}) we have the
expression
\begin{equation}
E_{L}[\mathbf{X}]=\mathbf{\dot{X}}\cdot \mathbf{P}^{0}[\mathbf{X}]+\mathbf{%
\ddot{X}}\cdot P^{1}[\mathbf{X}]-L[\mathbf{X}]  \label{EL}
\end{equation}%
for the energy function. The same idea works for Lagrangians of any finite
order.

Following \cite{run64}, let us show that the Lagrangians invariant under
reparametrization of curves $t\mapsto \mathbf{X}(t)$ are necessarily
degenerate and have zero energy. More precisely, we introduce new
parametrization $\tau =\tau (t)$, and let $\lambda \equiv dt/d\tau $, $\nu
\equiv d\lambda /d\tau =d^{2}t/d\tau ^{2}$. A second order Lagrangian is
reparametrization invariant if
\begin{equation}
\lambda L\mathbf{(X,\dot{X},\ddot{X})}=L(\mathbf{X},\lambda \mathbf{\dot{X}}%
,\lambda ^{2}\mathbf{\ddot{X}}+\nu \mathbf{\dot{X}})\text{,}  \label{rep}
\end{equation}%
\cite{DuSl09}. Since for $\lambda =1$ and $\nu =0$, we have equal
derivatives with respect to both parametrization, we differentiate above
equation with respect to $\lambda $ and $\nu $ at $(\lambda ,\nu )=(1,0)$ in
order to obtain the infinitesimal invariance conditions%
\begin{equation}
L=\mathbf{\dot{X}}\cdot {\frac{\partial L}{\partial \mathbf{\dot{X}}}}+2%
\mathbf{\ddot{X}}\cdot {\frac{\partial L}{\partial \mathbf{\ddot{X}}}}\text{%
, \ \ \ \ }\mathbf{\dot{X}}\cdot {\frac{\partial L}{\partial \mathbf{\ddot{X}%
}}}=0  \label{cond}
\end{equation}%
also known as Zermelo conditions \cite{UrKr13, Ze}. After solving $\partial
L/\partial \mathbf{\dot{X}}$ and $\partial L/\partial \mathbf{\ddot{X}}$ in
terms of momenta from Eq.(\ref{jo}), and substituting these into the first
condition in Eq.(\ref{cond}) we arrive that the Lagrangian must be in form
\begin{equation*}
L=\mathbf{\dot{X}}\cdot (\mathbf{P}^{0}+\mathbf{\dot{P}}^{1})+2\mathbf{\ddot{%
X}}\cdot \mathbf{P}^{1}=\mathbf{\dot{X}}\cdot \mathbf{P}^{0}+\mathbf{\ddot{X}%
}\cdot \mathbf{P}^{1},
\end{equation*}%
which results with that the energy function $E_{L}$ given in (\ref{EL}) is
zero. Differentiating the second condition in Eq.(\ref{cond}) with respect
to $\mathbf{\ddot{X}}$, we obtain a system of equations for $\mathbf{\dot{X}}
$ for which existence of non-zero solutions implies the degeneracy
\begin{equation*}
detHess(L)\equiv det\left[ \frac{\partial ^{2}L}{\partial \ddot{X}^{2}}%
\right] =0
\end{equation*}%
of the second order Lagrangian \cite{Ko91}. We remark here also that, from
the second condition in Eq.(\ref{cond}), we arrive a condition $\mathbf{\dot{%
X}}\cdot \mathbf{P}^{1}=0$ and its differential $\mathbf{\ddot{X}}\cdot
\mathbf{P}^{1}=-\mathbf{\dot{X}}\cdot \mathbf{\dot{P}}^{1}.$

\subsection{Dirac-Bergmann algorithm}

Consider a second order Lagrangian density $L=L[\mathbf{X}]$. As discussed
previously, the resulting Euler-Lagrange equations are singular, that is,
not all second derivatives are solvable, if the Hessian matrix $\partial
^{2}L/\partial \ddot{X}^{2}$ has rank $r<n$. That means there are only $n-r$
independent equations for derivatives higher than second order. The
Jacobi-Ostrogradsky momenta $\mathbf{P}^{0}$ and $\mathbf{P}^{1}$ become
functions of $\mathbf{(X,\dot{X},\ddot{X})}$ and $\mathbf{(X,\dot{X})}$,
respectively. More specifically, $\mathbf{P}^{0}$ is a linear function of $%
\ddot{\mathbf{X}}$ and the Euler-Lagrange equations are of third order. From
the definition of $\mathbf{P}^{1}$ we obtain relations%
\begin{equation*}
\Phi _{\alpha }(\mathbf{X,\dot{X}},\mathbf{P}^{1})\approx 0,\quad \alpha
=1,...,n-r
\end{equation*}%
among phase space coordinates, called primary constraints \cite%
{Dirac,Dirac1950}. From the equations defining the momenta $\mathbf{P}^{0}$,
only some of the second order derivatives are solvable. It is also possible
to define constraints from definition of $\mathbf{P}^{0}$ if there are
equations independent of second derivatives. However, as it is shown in \cite%
{SSOK} all such possible constraints can be derived from the preservation of
primary constraints introduced immediately$.$

In fact, for any Lagrangian in which the second derivative term is expressed
as a triple product (more generally, is involved in completely antisymmetric
tensor), only two components of second derivatives can be solved from
definition of momenta $\mathbf{P}^{0}$, the last one is satisfied
identically. This dimensional degeneracy is the reason behind definition of
some constraints by means of dot product.

The equality in the definition of primary constraints is weak in the sense
that it is ignored during set up of Dirac formalism, and will actually
vanish in any solutions to equations of motion. In other words, $\Phi
_{\alpha }$ are not identically zero on the phase space but vanish on
primary constraint submanifold they define. The dynamics on this submanifold
is not well-defined by the canonical Hamiltonian function%
\begin{equation*}
H=\mathbf{P}^{0}\cdot \mathbf{\dot{X}}+\mathbf{P}^{1}\cdot \mathbf{\ddot{X}}%
-L[\mathbf{X}],
\end{equation*}%
it is rather governed by the total Hamiltonian
\begin{equation*}
H_{T}=H+u^{\alpha }\Phi _{\alpha }
\end{equation*}%
which contains linear combinations of primary constraints with Lagrange
multipliers $u^{\alpha }$. The requirement that the solutions of
Euler-Lagrange equations remain on constraint submanifold is described by
the weak equality%
\begin{equation}
\dot{\Phi}_{\beta }=\{\Phi _{\beta },H\}+u^{\alpha }\{\Phi _{\beta },\Phi
_{\alpha }\}\approx 0\text{, \ \ \ }\beta =1,...,n-r,
\end{equation}%
that is, modulo primary constraints. These consistency conditions may lead
to determination of Lagrange multipliers if the left hand sides contain $%
u^{\alpha }$. In this case, one solves for $u^{\alpha }$ through the set of
linear equations%
\begin{equation*}
\{\Phi _{\beta },\Phi _{\alpha }\}u^{\alpha }=-\{\Phi _{\beta },H\}
\end{equation*}%
for which the solution set, namely, number of multipliers that can be solved
is characterized by the rank of the skew-symmetric matrix $\{\Phi _{\beta
},\Phi _{\alpha }\}$ of Poisson brackets. Obviously, if the number of
primary constraints is odd $u^{\alpha }$s cannot be solved completely and
one aspects more constraints to determine $H_{T}$ in terms of phase space
variables. This secondary constraints follow if left hand sides does not
contain $u^{\alpha }$ or, $n-r$ is odd. Repeating this process, one enlarges
the primary constraint set with the new (secondary, tertiary, etc.)
constraints, redefines $H_{T}$ by introducing new Lagrange multipliers for
new constraints and, repeats the consistency computations.

Iterated applications of consistency computations lead to a complete set of
constraints $\Phi _{\alpha }:\alpha =1,...,k$. Let
\begin{equation*}
\mathcal{M}_{\alpha \beta }=\left\{ \Phi _{\alpha },\Phi _{\beta }\right\}
\end{equation*}%
be the matrix of Poisson brackets of constraints modulo all constraints. If $%
rank(M_{\alpha \beta })=r$, then $ker(M_{\alpha \beta })$ is $\left(
k-r\right) -$dimensional. A basis for the kernel can be constructed from
linear combinations $\psi _{\alpha }$ of $\Phi _{\alpha }$ satisfying%
\begin{equation*}
\left\{ \psi _{\alpha },\psi _{\beta }\right\} \approx 0,\text{ \ }\alpha
,\beta =1,...,k-r
\end{equation*}%
and are called first class constraints. Note that the number of Lagrange
multipliers which can be solved is also determined by the matrix of all
constraints. Let $\chi _{\alpha }:\alpha =1,...,r$ be the second class
constraints whose Poisson brackets does not vanish (modulo constraints).
Define the $r\times r-$matrix%
\begin{equation*}
C_{\alpha \beta }=\left\{ \chi _{\alpha },\chi _{\beta }\right\} ,\text{ \ }%
\alpha ,\beta =1,...,r
\end{equation*}%
which is invertible by construction. Define the Dirac bracket%
\begin{equation}
\{f,g\}_{DB}=\{f,g\}-\{f,\chi _{\alpha }\}(C^{-1})^{\alpha \beta }\{\chi
_{\beta },g\}  \label{diracbrac}
\end{equation}%
\cite{SuMu74}. Note that, since $\{f,\chi _{\alpha }\}_{DB}=0$ for arbitrary
function $f$, second class constraints can be set to zero either before or
after evaluation of Dirac bracket. The initial $2n-$dimensional Hamiltonian
system with $k-r-$first class and $r-$second class constraints becomes $%
2n-2(k-r)-r=2n-2k+r-$dimensional reduced Hamiltonian system for the Dirac
bracket and with the total Hamiltonian function of Dirac. The final bracket
eliminates the second class constraints from the set of all constraints
leaving a complete set of first class constraints. First class constraints
form a closed local symmetry algebra for the system. Computing%
\begin{equation*}
\left\{ \psi _{\alpha },H\right\} =c_{\alpha }^{\beta }\psi _{\beta },\text{
\ \ }\left\{ \psi _{\alpha },\psi _{\beta }\right\} =c_{\alpha \beta
}^{\gamma }\psi _{\gamma }
\end{equation*}%
one finds the structure constants of this algebra \cite{Dirac}, \cite%
{Dirac1950}.

\subsection{Skinner-Rusk unified formalism}

The Skinner-Rusk unified formalism is to define a proper submanifold of the
presymplectic Pontryagin bundle, Whitney product of velocity and momentum
phase spaces, which enables one to study the Hamiltonian and Lagrangian
formalism altogether \cite{Sk83,SkRu83,SkRu83b}. By following \cite{PrRo11},
let us summarize the Skinner-Rusk unified formalism in the case of second
order Lagrangians \cite{PrRo11}. We refer \cite{PrRo12} for the
non-autonomous cases.

Consider the second order Pontryagin bundle
\begin{equation}
P^{3}Q=T^{3}Q\times _{TQ}T^{\ast }TQ  \label{P3M}
\end{equation}%
which is the Whitney product of the third order tangent bundle $T^{3}Q$ and
the iterated cotangent bundle $T^{\ast }TQ$ over the base manifold $TQ$. The
induced coordinates on $P^{3}Q$ is given by six-tuples
\begin{equation}
\mathbf{(X,\dot{X},\ddot{X},\dddot{X},P}^{0}\mathbf{,P}^{1}\mathbf{)}\in
P^{3}Q  \label{locP}
\end{equation}%
obtained those defined on $T^{3}Q$ and $T^{\ast }TQ$. There are projections $%
pr_{1}$ and $pr_{2}$ from $P^{3}Q$ to $T^{3}Q$ and $T^{\ast }TQ$,
respectively. Skinner-Rusk formalism on the second order bundle is to search
a possible solution of the presymplectic Hamilton's equation
\begin{equation}
i_{X_{P^{3}Q}}\Omega _{P^{3}Q}=-dE,  \label{PreHam2}
\end{equation}%
where the Hamiltonian function $E$ is assumed to be the energy function (\ref%
{EL}) in form
\begin{equation}
E\mathbf{(X,\dot{X},\ddot{X},\dddot{X},P}^{0}\mathbf{,P}^{1}\mathbf{)}=%
\mathbf{P}^{0}\cdot \mathbf{\dot{X}+P}^{1}\cdot \mathbf{\ddot{X}-L}\left(
\mathbf{X,\dot{X},\ddot{X}}\right) .  \label{canHam}
\end{equation}
Here, $\Omega _{P^{3}Q}$ is the presymplectic two-form on $P^{3}Q$ and
obtained by pull-back of the canonical symplectic two-form $\Omega _{T^{\ast
}TQ}$ on $T^{\ast }TQ$ by the projection $pr_{2}$.

In the local chart (\ref{locP}), $\Omega _{P^{3}Q}$ is computed to be
\begin{equation}
\Omega _{P^{3}Q}=\left( pr_{2}\right) ^{\ast }\Omega _{T^{\ast }TQ}=d\mathbf{%
P}^{0}\wedge d\mathbf{X}+d\mathbf{P}^{1}\wedge d\mathbf{\dot{X},}
\label{OhmP3M}
\end{equation}%
whereas the Hamiltonian vector field looks like%
\begin{equation}
X_{P^{3}M}=\mathbf{\dot{X}}\cdot \nabla _{\mathbf{X}}+\mathbf{\ddot{X}}\cdot
\nabla _{\mathbf{\dot{X}}}+\mathbf{\dddot{X}}\cdot \nabla _{\mathbf{\ddot{X}}%
}+\mathbf{B}\cdot \nabla _{\mathbf{\dddot{X}}}+\nabla _{\mathbf{X}}L\cdot
\nabla _{\mathbf{P}^{0}}+\left( \nabla _{\mathbf{\dot{X}}}L-\mathbf{P}%
^{0}\right) \cdot \nabla _{\mathbf{P}^{1}}  \label{X_P3M}
\end{equation}%
with compatibility conditions $\mathbf{P}^{1}=\nabla _{\mathbf{\ddot{X}}}L$
to be sure that $X_{P^{3}M}$ is tangent to $W_{0}$ (c.f. the second Jacobi
Ostrogradsky momenta (\ref{jo})). This assumption is necessary to guarantee
that the projection
\begin{equation*}
X_{T^{3}M}=\left( pr_{1}\right) _{\ast }X_{P^{3}M}
\end{equation*}%
is a Euler-Lagrange vector field, that is the Hamilton's equations
\begin{equation*}
i_{X_{T^{3}M}}\Omega _{T^{3}M}=-dE_{L}
\end{equation*}%
give Euler-Lagrange equations on\ the submanifold $pr_{1}\left( W_{f}\right)
$ of $T^{3}M$.

Finding a vector field $X_{P^{3}M}$ satisfying Hamilton's equations (\ref%
{PreHam2}) is possible on a submanifold $W_{f}$, so called final constraint
submanifold, of $P^{3}M$. We start with determining the primary constraint
submanifold $W_{0}$ by defining the primary constraints
\begin{equation}
\mathbf{\Psi }=\mathbf{P}^{0}-\nabla _{\mathbf{\dot{X}}}L+\frac{d}{dt}\nabla
_{\mathbf{\ddot{X}}}L,\text{ \ \ }\mathbf{\Phi }=\mathbf{P}^{1}-\nabla _{%
\mathbf{\ddot{X}}}L.
\end{equation}%
If the tangency conditions
\begin{equation}
X_{P^{3}M}\left( \mathbf{\Psi }\right) =\mathbf{0}\text{ \ \ and \ \ }%
X_{P^{3}M}\left( \mathbf{\Phi }\right) =\mathbf{0}  \label{tangcond}
\end{equation}%
hold, then the final constraint submanifold $W_{f}$ equals to the primary
constraint submanifold $W_{0}$. This occurs if the Lagrangian is regular,
that is the tangent map of the Jacobi-Ostrogradsky momenta (\ref{jo}) is
surjective submersion at every point in its domain. If the Lagrangian is
degenerate, then the tangency conditions (\ref{tangcond}) lead to two new
sets of constraints%
\begin{equation*}
\mathbf{\Psi }_{1}=X_{P^{3}M}\mathbf{\Psi }\text{, \ \ }\mathbf{\Phi }%
_{2}=X_{P^{3}M}\mathbf{\Phi }.
\end{equation*}%
and, correspondingly, a constraint submanifold $W_{1}$ of $W_{0}$ by
additionally requiring $\mathbf{\Psi }_{1}=\mathbf{\Phi }_{1}=\mathbf{0}$.
If we ask the tangency conditions
\begin{equation*}
X_{P^{3}M}\left( \mathbf{\Psi }_{1}\right) =\mathbf{0}\text{ \ \ and \ \ }%
X_{P^{3}M}\left( \mathbf{\Phi }_{1}\right) =\mathbf{0}
\end{equation*}%
for the constraints then there are possible scenarios. The first one is to
observe that $X_{P^{3}M}\mathbf{\Psi }_{1}$ and $X_{P^{3}M}\mathbf{\Phi }%
_{1} $ are identically zero. This gives that $W_{1}$ is the final
submanifold $W_{f}$ and we are done. The second one is to arrive at two new
set of constraints%
\begin{equation*}
\mathbf{\Psi }_{2}=X_{P^{3}M}\left( \mathbf{\Psi }_{1}\right) \text{, \ \ }%
\mathbf{\Phi }_{2}=X_{P^{3}M}\mathbf{\Phi }_{1},
\end{equation*}%
called the first-generation secondary constraints. Using these constraints,
define a submanifold $W_{2}$ of $W_{1}$ by additionally requiring $\mathbf{%
\Psi }_{2}=\mathbf{\Phi }_{2}=0$. Repeating this algorithm, we may obtain $%
k- $generation secondary constraints which defines a submanifold $W_{k}$. If
the nested sequence
\begin{equation*}
W_{k}\subset W_{k-1}\subset ...\subset W_{0}
\end{equation*}%
has an end, that is $W_{k+1}=W_{k}$, then $W_{k}$ is the final constraint
submanifold and on this submanifold and the vector field $X_{P^{3}M}$ has a
well-defined expression satisfying Eq.(\ref{PreHam2}).

\section{Hamiltonian analysis of Sar\i o\u{g}lu-Tekin Lagrangian}

\subsection{Sar\i o\u{g}lu-Tekin Lagrangian}

We start with a $6$-dimensinal manifold $Q$ with local coordinates $(\mathbf{%
X,Y})$ consisting of two $3$-dimensional vectors. The higher order tangent
bundles are equipped with the following induced sets of coordinates
\begin{eqnarray*}
(\mathbf{X,Y,\dot{X},\dot{Y}}) &\in &TQ \\
(\mathbf{X,Y,\dot{X},\dot{Y},\ddot{X},\ddot{Y}}) &\in &T^{2}Q \\
(\mathbf{X,Y,\dot{X},\dot{Y},\ddot{X},\ddot{Y},\dddot{X},\dddot{Y}}) &\in
&T^{3}Q.
\end{eqnarray*}%
In \cite{st06}, Sar\i o\u{g}lu and Tekin proposed a degenerate second order
Lagrangian on $T^{2}Q$ given by%
\begin{equation}
L^{ST}[\mathbf{X},\mathbf{Y}]=\frac{1}{2}\left[ a(\dot{X}^{2}+\dot{Y}^{2})+%
\frac{2}{\mu }\dot{\mathbf{Y}}\cdot \ddot{\mathbf{X}}-m^{2}(Y^{2}+X^{2})%
\right] .  \label{LasST}
\end{equation}%
In this case, the second order Euler-Lagrange equations (\ref{ele}) take the
particular form
\begin{equation}
m^{2}\mathbf{X}+a\mathbf{\ddot{X}}=\frac{1}{\mu }\mathbf{Y}^{(3)},\text{\ \
\ }m^{2}\mathbf{Y}+a\mathbf{\ddot{Y}}=-\frac{1}{\mu }\mathbf{X}^{(3)}.
\label{ste}
\end{equation}%
The Lagrangian one-form (\ref{theta}) takes the particular form
\begin{equation*}
\theta _{L}=(a\mathbf{\dot{X}}-\frac{1}{\mu }\mathbf{\ddot{Y}})\cdot d%
\mathbf{X}+(a\mathbf{\dot{Y}}+\frac{1}{\mu }\mathbf{\ddot{X}})\cdot d\mathbf{%
Y}+\frac{1}{\mu }\mathbf{\dot{Y}}\cdot d\mathbf{\dot{X},}
\end{equation*}%
on the second order tangent bundle $T^{2}Q$ whereas the exterior derivative
of $\theta _{L}$ becomes%
\begin{equation*}
\Omega _{L}=a(d\mathbf{\dot{X}}\dot{\wedge}d\mathbf{X+}d\mathbf{\dot{Y}}\dot{%
\wedge}d\mathbf{Y})+\frac{1}{\mu }d\mathbf{\dot{Y}}\dot{\wedge}d\mathbf{\dot{%
X}+}\frac{1}{\mu }(d\mathbf{\ddot{X}}\dot{\wedge}d\mathbf{Y}-d\mathbf{\ddot{Y%
}}\dot{\wedge}d\mathbf{X}).
\end{equation*}%
Here, we use the abbreviation $\dot{\wedge}$ defined as
\begin{equation}
d\mathbf{X}\dot{\wedge}d\mathbf{Y}=dX^{1}\wedge dY^{1}+dX^{2}\wedge
dY^{2}+dX^{3}\wedge dY^{3}.  \label{wedgedot}
\end{equation}%
In this notation, $d\mathbf{X}\dot{\wedge}d\mathbf{X}=0$ identically.

The Sar\i o\u{g}lu-Tekin Lagrangian (\ref{LasST}) has $SO(3)$ invariance
resulting in the momenta%
\begin{equation*}
\frac{1}{2}J[\mathbf{X},\mathbf{Y}]=a\mathbf{Y}\times \mathbf{\dot{Y}}+\frac{%
1}{\mu }\mathbf{Y}\times \mathbf{\ddot{X}}+a\mathbf{X}\times \mathbf{\dot{X}}%
-\frac{1}{\mu }\mathbf{X}\times \mathbf{\ddot{Y}}+\frac{1}{\mu }\mathbf{\dot{%
X}}\times \mathbf{\dot{Y}}
\end{equation*}%
and the time-translation invariance gives the energy%
\begin{equation*}
E^{ST}[\mathbf{X},\mathbf{Y}]=a(\dot{X}^{2}+\dot{Y}^{2})+\frac{2}{\mu }(%
\mathbf{\dot{Y}\cdot \ddot{X}}-\mathbf{\dot{X}}\cdot \mathbf{\ddot{Y}}%
)+m^{2}(Y^{2}+X^{2})
\end{equation*}%
both of which may be shown to satisfy the conservation laws $\dot{J}=\dot{E}%
^{ST}=0$ via Euler-Lagrange equations.

\subsection{Dirac-Bergmann Algorithm}

The iterated cotangent bundle $T^{\ast }TQ$ is $24-$dimensional and equipped
with a local chart
\begin{equation*}
(\mathbf{X,Y,\dot{X},\dot{Y},P}_{X}^{0},\mathbf{P}_{Y}^{0},\mathbf{P}%
_{X}^{1},\mathbf{P}_{Y}^{1})\in T^{\ast }TQ.
\end{equation*}%
The Jacobi-Ostrogradsky momenta (\ref{jo}) are defined by
\begin{equation}
\mathbf{P}_{X}^{0}=a\mathbf{\dot{X}}-\frac{1}{\mu }\mathbf{\ddot{Y}},\;\;%
\mathbf{P}_{X}^{1}=\frac{1}{\mu }\mathbf{\dot{Y}}\ \ \ \mathbf{P}_{Y}^{0}=a%
\mathbf{\dot{Y}}+\frac{1}{\mu }\mathbf{\ddot{X}},\;\;\mathbf{P}_{Y}^{1}=%
\mathbf{0},  \label{OstmomST}
\end{equation}%
whereas the canonical Hamiltonian functions is%
\begin{equation}
H^{ST}=\mathbf{P}_{X}^{0}\cdot \dot{\mathbf{X}}+\mathbf{P}_{Y}^{0}\cdot \dot{%
\mathbf{Y}}-\frac{a}{2}(\dot{X}^{2}+\dot{Y}^{2})+\frac{m^{2}}{2}%
(X^{2}+Y^{2}).  \label{HamcanST}
\end{equation}

The momenta (\ref{OstmomST}) give the following two second derivatives
\begin{equation}
\mathbf{\ddot{Y}}=\mu (a\mathbf{\dot{X}}-\mathbf{P}_{X}^{0}),\ \ \ \mathbf{%
\ddot{X}}=\mu (\mathbf{P}_{Y}^{0}-a\mathbf{\dot{Y})},\;  \label{2ndST}
\end{equation}%
along with and the set of primary constraints%
\begin{equation}
\mathbf{\Phi =P}_{X}^{1}-\frac{1}{\mu }\mathbf{\dot{Y}\approx 0}\text{, \ \
\ \ }\;\mathbf{\Psi =P}_{Y}^{1}\approx \mathbf{0.}  \label{primST}
\end{equation}%
We define the total Hamiltonian
\begin{equation*}
H_{T}^{ST}=H^{ST}+\mathbf{U}\cdot \mathbf{\Phi }+\mathbf{V}\cdot \mathbf{%
\Psi }
\end{equation*}%
as the sum of the canonical Hamiltonian $H^{ST}$ in (\ref{HamcanST}) and the
primary constraints $\mathbf{\Phi }$ and $\mathbf{\Psi }$ in (\ref{primST})
multiplied by the Lagrange multipliers $\mathbf{U}$ and $\mathbf{V}$,
respectively. The consistency check results with determination of the
Lagrange multipliers%
\begin{equation}
\mathbf{V}=\mu (a\mathbf{\dot{X}}-\mathbf{P}_{X}^{0})\text{\textbf{, \ \ }}%
\mathbf{U}=\mu (\mathbf{P}_{Y}^{0}-a\mathbf{\dot{Y})}\text{\textbf{\ }}
\label{LmST}
\end{equation}%
without causing a new constraint. This means that we have arrived the final
submanifold.

After the substitution of the Lagrange multipliers $\mathbf{V}$ and $\mathbf{%
U}$ in (\ref{LmST}) into the total Hamiltonian $H_{T}^{ST}$, the total turns
out to be
\begin{eqnarray}
H_{T}^{ST} &=&\mu (\mathbf{P}_{Y}^{0}\cdot \mathbf{P}_{X}^{1}-\mathbf{P}%
_{X}^{0}\cdot \mathbf{P}_{Y}^{1})+a\mu (\mathbf{\dot{X}}\cdot \mathbf{P}%
_{Y}^{1}-\mathbf{\dot{Y}}\cdot \mathbf{P}_{X}^{1})+\mathbf{P}_{X}^{0}\cdot
\mathbf{\dot{X}}  \notag \\
&&-\frac{a}{2}(\dot{X}^{2}-\dot{Y}^{2})+\frac{m^{2}}{2}(X^{2}+Y^{2}).
\label{TH-SK}
\end{eqnarray}%
Note that, to arrive at the total Hamiltonian function (\ref{TH-SK}) , we
may follow a more direct way by solving $\mathbf{\ddot{X}}$ and $\mathbf{%
\ddot{Y}}$ from the equations (\ref{OstmomST}) and substituting $\mathbf{%
\ddot{X}}$ and $\mathbf{\ddot{Y}}$ in Eqs.(\ref{OstmomST}) into the
canonical Hamiltonian function without referring any constraint analysis.
For the base variables $\left( \mathbf{X,Y,\dot{X},\dot{Y}}\right) ,$ the
equations of motion governed by the total Hamiltonian $H_{T}^{ST}$ are%
\begin{equation}
\mathbf{\dot{X}}=\mathbf{\dot{X}}-\mu \mathbf{P}_{Y}^{1}\text{, \ \ }\mathbf{%
\ddot{X}}=\mu \mathbf{P}_{Y}^{0}-a\mu \mathbf{\dot{Y}}\text{, \ \ }\mathbf{%
\dot{Y}}=\mu \mathbf{P}_{X}^{1}\text{, \ \ }\mathbf{\ddot{Y}}=-\mu \mathbf{P}%
_{X}^{0}+a\mu \mathbf{\dot{X}}  \label{HamEqST1}
\end{equation}%
which are satisfied identically on the constraint submanifold. For momenta $%
\left( \mathbf{P}_{X}^{0},\mathbf{P}_{Y}^{0},\mathbf{P}_{X}^{1},\mathbf{P}%
_{Y}^{1}\right) $, the equations of motion are
\begin{equation}
\mathbf{\dot{P}}_{X}^{0}=-m^{2}\mathbf{X}\text{, \ \ }\mathbf{\dot{P}}%
_{X}^{1}=a\mathbf{\dot{X}}-a\mu \mathbf{P}_{Y}^{1}-\mathbf{P}_{X}^{0}\text{,
\ \ }\mathbf{\dot{P}}_{Y}^{0}=-m^{2}\mathbf{Y}\text{, \ \ }\mathbf{\dot{P}}%
_{Y}^{1}=-a\mathbf{\dot{Y}}+a\mu \mathbf{P}_{X}^{1},  \label{HamEqST2}
\end{equation}%
where the second and the fourth ones are identically satisfied. The first
and third equations give Euler-Lagrange equations (\ref{ste}) only after the
substitution of the second order equations (\ref{2ndST}).

\subsection{Dirac-Poisson Bracket}

The constraints $\mathbf{\Phi }$ and$\;\mathbf{\Psi }$ are all second class,
hence the Poisson brackets of them define the nondegenerate $6\times 6$
constraint matrix%
\begin{equation*}
C=\left(
\begin{array}{cc}
\{\Phi _{i},\Phi _{j}\} & \{\Phi _{i},\Psi _{j}\} \\
\{\Psi _{j},\Phi _{i}\} & \{\Psi _{i},\Psi _{j}\}%
\end{array}%
\right) =\frac{1}{\mu }\left(
\begin{array}{cc}
\mathbf{0} & -\mathbb{I} \\
\mathbb{I} & \mathbf{0}%
\end{array}%
\right) .
\end{equation*}%
Hence, the Dirac bracket in Eq.(\ref{diracbrac}) takes the particular form
\begin{equation}
\{F,G\}_{DB}=\{F,G\}-\{F,\Phi _{i}\}\mu \delta ^{ij}\{\Psi _{j},G\}+\{F,\Psi
_{i}\}\mu \delta ^{ij}\{\Phi _{j},G\}  \label{diracbracforsar}
\end{equation}%
and results in the reduced Poisson structure
\begin{equation}
\{X^{i},(P_{X}^{0})_{j}\}_{DB}=\{\dot{X}^{i},(P_{X}^{1})_{j}\}_{DB}=%
\{Y^{i},(P_{Y}^{0})_{j}\}_{DB}=\delta _{j}^{i}\text{, \ \ }\{\dot{X}^{i},%
\dot{Y}^{j}\}_{DB}=\mu \delta ^{ij}\text{.}  \label{DB-SK}
\end{equation}

It is straight-forward to check that, using the Dirac bracket (\ref%
{diracbracforsar}), the equations of motion generated by the canonical
Hamiltonian $H^{ST}$ in Eq.(\ref{HamcanST}) is exactly equal to the dynamics
generated by the $H_{T}^{ST}$.

\subsection{Skinner-Rusk unified formalism}

On the Pontryagin bundle $P^{3}Q=T^{3}Q\times _{TQ}T^{\ast }TQ$, the
presymplectic two-from $\Omega _{P^{3}Q}$ defined in Eq.(\ref{OhmP3M}), and
the canonical Hamiltonian function defined in (\ref{canHam}) turn out to be%
\begin{equation}
\Omega _{P^{3}Q}=pr_{2}^{\ast }\Omega _{T^{\ast }TQ}=d\mathbf{P}%
_{X}^{0}\wedge d\mathbf{X}+d\mathbf{P}_{Y}^{0}\wedge d\mathbf{Y}+d\mathbf{P}%
_{X}^{1}\wedge d\mathbf{\dot{X}}+d\mathbf{P}_{Y}^{1}\wedge d\mathbf{\dot{Y}}
\label{OhmP}
\end{equation}%
whereas the canonical Hamiltonian function%
\begin{equation*}
H_{P^{3}Q}=\mathbf{P}_{X}^{0}\cdot \mathbf{\dot{X}}+\mathbf{P}_{Y}^{0}\cdot
\mathbf{\dot{Y}}+\mathbf{P}_{X}^{1}\cdot \mathbf{\ddot{X}}+\mathbf{P}%
_{Y}^{1}\cdot \mathbf{\ddot{Y}}-L^{ST}[\mathbf{X},\mathbf{Y}].
\end{equation*}

To determine a unique vector field $X_{P^{3}Q}$ satisfying the Hamilton's
equations (\ref{PreHam2}), we start with%
\begin{eqnarray}
X_{P^{3}Q} &=&\mathbf{\dot{X}}\cdot \nabla _{\mathbf{X}}+\mathbf{\ddot{X}}%
\cdot \nabla _{\mathbf{\dot{X}}}+\mathbf{\dddot{X}}\cdot \nabla _{\mathbf{%
\ddot{X}}}+\mathbf{K}_{X}\cdot \nabla _{\mathbf{\dddot{X}}}+\mathbf{\dot{Y}}%
\cdot \nabla _{\mathbf{Y}}+\mathbf{\ddot{Y}}\cdot \nabla _{\mathbf{\dot{Y}}}+%
\mathbf{\dddot{Y}}\cdot \nabla _{\mathbf{\ddot{Y}}}  \notag \\
&&+\mathbf{K}_{Y}\cdot \nabla _{\mathbf{\dddot{Y}}}-m^{2}\mathbf{X}\cdot
\nabla _{\mathbf{P}_{X}^{0}}-m^{2}\mathbf{Y}\cdot \nabla _{\mathbf{P}%
_{Y}^{0}}+\left( a\mathbf{\dot{X}-P}_{X}^{0}\right) \cdot \nabla _{\mathbf{P}%
_{X}^{1}}  \notag \\
&&+\left( a\mathbf{\dot{Y}-P}_{Y}^{0}+\frac{1}{\mu }\mathbf{\ddot{X}}\right)
\cdot \nabla _{\mathbf{P}_{Y}^{1}},  \label{vf-SR-ST}
\end{eqnarray}%
where we have two sets of unknown coefficient functions $\mathbf{K}_{X}$ and
$\mathbf{K}_{Y}$. The graph of the Legendre transformation $\mathcal{F}%
L^{ST} $ is described by the following primary constraints%
\begin{eqnarray}
\mathbf{\bar{\Phi}} &=&\mathbf{P}_{X}^{0}-a\mathbf{\dot{X}+}\frac{1}{\mu }%
\mathbf{\ddot{Y},}\text{ \ \ }\mathbf{\bar{\Psi}}=\mathbf{P}_{Y}^{0}-a%
\mathbf{\dot{Y}+}\frac{1}{\mu }\mathbf{\ddot{X}}  \notag \\
\mathbf{\Phi } &=&\mathbf{P}_{X}^{1}-\frac{1}{\mu }\mathbf{\dot{Y}}\text{, \
\ }\mathbf{\Psi }=\mathbf{P}_{Y}^{1}.  \notag
\end{eqnarray}%
The first step is to check the tangency conditions:%
\begin{eqnarray}
X_{P^{3}N}\mathbf{\bar{\Phi}} &=&\mathbf{\bar{\Phi}}_{1}=\frac{1}{\mu }%
\mathbf{\dddot{Y}}-m^{2}\mathbf{X}-a\mathbf{\ddot{X},}  \notag \\
X_{P^{3}N}\mathbf{\bar{\Psi}} &=&\mathbf{\bar{\Psi}}_{1}=\frac{1}{\mu }%
\mathbf{\dddot{X}+}m^{2}\mathbf{Y+a\ddot{Y},}  \notag \\
X_{P^{3}N}\mathbf{\Phi } &=&-\mathbf{\bar{\Phi}},  \notag \\
X_{P^{3}N}\mathbf{\Psi } &=&-\mathbf{\bar{\Psi}.}
\end{eqnarray}%
The last two equations are weakly zero, so that we only take the first two $%
\mathbf{\bar{\Phi}}_{1}\ $and $\mathbf{\bar{\Psi}}_{1}$ as new constraints.
The first constraint submanifold $W_{1}$ is defined by the set of functions $%
\mathbf{\bar{\Phi}},\mathbf{\bar{\Psi}},\mathbf{\Phi },\mathbf{\Psi ,\mathbf{%
\bar{\Phi}}_{1},\bar{\Psi}}_{1}$. For the secondary constraints, we have
that
\begin{eqnarray*}
X_{P^{3}N}\mathbf{\mathbf{\bar{\Phi}}_{1}} &=&\frac{1}{\mu }\mathbf{K}%
_{Y}-m^{2}\mathbf{\dot{X}}-a\mathbf{\dddot{X}} \\
X_{P^{3}N}\mathbf{\bar{\Psi}}_{1} &=&-\frac{1}{\mu }\mathbf{K}_{X}-m^{2}%
\mathbf{\dot{Y}}-a\mathbf{\dddot{Y}}
\end{eqnarray*}%
from which, by requiring that they weakly equal to zero, we obtain unknown
functions
\begin{eqnarray*}
\mathbf{K}_{Y} &=&\mu m^{2}\mathbf{\dot{X}}+\mu a\mathbf{\dddot{X}} \\
\mathbf{K}_{X} &=&-\mu m^{2}\mathbf{\dot{Y}}-a\mu \mathbf{\dddot{Y}}.
\end{eqnarray*}%
We have no secondary constraints and that $W_{1}=W_{f}$ is the final
constraint submanifold.

We obtain the Euler-Lagrange vector field $X_{T^{3}Q}$ on $T^{3}Q$ by
projecting $X_{P^{3}Q}$ in (\ref{vf-SR-ST}) via $pr_{1}$, that is
\begin{eqnarray}
X_{T^{3}N} &=&\mathbf{\dot{X}}\cdot \nabla _{\mathbf{X}}+\mathbf{\ddot{X}}%
\cdot \nabla _{\mathbf{\dot{X}}}+\mathbf{\dddot{X}}\cdot \nabla _{\mathbf{%
\ddot{X}}}-\left( \mu m^{2}\mathbf{\dot{Y}}+a\mu \mathbf{\dddot{Y}}\right)
\cdot \nabla _{\mathbf{\dddot{X}}}+  \notag \\
&&+\mathbf{\dot{Y}}\cdot \nabla _{\mathbf{Y}}+\mathbf{\ddot{Y}}\cdot \nabla
_{\mathbf{\dot{Y}}}+\mathbf{\dddot{Y}}\cdot \nabla _{\mathbf{\ddot{Y}}%
}+\left( \mu m^{2}\mathbf{\dot{X}}+\mu a\mathbf{\dddot{X}}\right) \cdot
\nabla _{\mathbf{\dddot{Y}}}.  \label{X-SK}
\end{eqnarray}%
The energy $E_{T^{3}Q}=\left( pr_{1}\right) ^{\ast }E_{P^{3}Q}$ is given by
\begin{equation}
E_{T^{3}Q}=\left( a\mathbf{\dot{X}}-\frac{1}{\mu }\mathbf{\ddot{Y}}\right)
\cdot \mathbf{\dot{X}}+\left( a\mathbf{\dot{Y}}+\frac{2}{\mu }\mathbf{\ddot{X%
}}\right) \cdot \mathbf{\dot{Y}}-L^{ST},  \label{E-SK}
\end{equation}%
and it satisfies the Hamilton's equations
\begin{equation*}
i_{X_{T^{3}Q}}\Omega _{T^{3}Q}=-dE_{T^{3}Q}
\end{equation*}%
on $S_{f}=pr^{1}\left( W_{f}\right) $ if the Euler-Lagrange equations (\ref%
{ste}) hold. Here, $\Omega _{T^{3}Q}$ is the two-form obtained by the push
forward of $\Omega _{P^{3}Q}$ in Eq.(\ref{OhmP}) to $T^{3}Q$.

\section{Hamiltonian analysis of Cl\`{e}ment Lagrangian \label{Clement-Lag}}

\subsection{Cl\`{e}ment Lagrangian}

Let us record here Cl\`{e}ment's degenerate second order Lagrangian density
\begin{equation}
L^{C}[\mathbf{X}]=-\frac{m}{2}\zeta \dot{X}^{2}-\frac{2m\Lambda }{\zeta }+%
\frac{\zeta ^{2}}{2\mu m}\mathbf{X}\cdot (\mathbf{\dot{X}}\times \mathbf{%
\ddot{X}})  \label{LC}
\end{equation}%
on the second order tangent bundle $T^{2}M$ with local coordinates $[\mathbf{%
X}]=(\mathbf{X,\dot{X},\ddot{X})}$. Here, the inner product $%
X^{2}=T^{2}-X^{2}-Y^{2}$ is defined by the Lorentzian metric and the triple
product is $\mathbf{X}\cdot (\mathbf{\dot{X}}\times \mathbf{\ddot{X}}%
)=\epsilon _{ijk}X^{i}\dot{X}^{j}\ddot{X}^{k}$ where $\epsilon _{ijk}$ is
the completely antisymmetric tensor of rank three. Dot denotes the
derivative with respect to the variable $t$ and $\zeta =\zeta (t)$ is a
function which allows arbitrary reparametrization of the variable $t$. $%
\Lambda $ and $1/2m$ are cosmological and Einstein gravitational constants,
respectively.

The variation of Cl\`{e}ment Lagrangian (\ref{LC}) with respect to $\zeta $
gives the energy constraint%
\begin{equation}
E^{C}[\mathbf{X}]=-\frac{m}{2}\dot{X}^{2}+2\frac{m\Lambda }{\zeta ^{2}}+%
\frac{\zeta }{m\mu }\mathbf{X}\cdot (\mathbf{\dot{X}}\times \mathbf{\ddot{X}}%
)=0.  \label{lcon}
\end{equation}%
whereas the variation of the Lagrangian (\ref{LC}) with respect to $\mathbf{X%
}$ results with the third order Euler-Lagrange equations
\begin{equation}
2m^{2}\mu \mathbf{\ddot{X}}+3\mathbf{\dot{X}}\times \mathbf{\ddot{X}}+2%
\mathbf{X}\times \mathbf{X}^{(3)}=\mathbf{0}.  \label{clee}
\end{equation}%
In the Euler-Lagrange equations (\ref{clee}), we set the reparametrization
function $\zeta $ equal to one. The Cl\`{e}ment Lagrangian (\ref{LC}) is
invariant under translations in $t$ and pseudo-rotations in space. Time
translation symmetry gives the conservation of energy in Eq.(\ref{lcon}).

Following the definitions (\ref{theta}) and (\ref{omega}), we compute the
Lagrange one-form $\theta _{L}[\mathbf{X}]$ and the pre-symplectic two-form $%
\Omega _{L}[\mathbf{X}]$ on $T^{2}M$ as follows
\begin{eqnarray}
\theta _{L}[\mathbf{X}] &=&-(m\zeta \mathbf{\dot{X}}+{\frac{\zeta ^{2}}{\mu m%
}}\mathbf{X}\times \mathbf{\ddot{X}})\cdot d\mathbf{X}+{\frac{\zeta ^{2}}{%
2\mu m}}(\mathbf{X}\times \mathbf{\dot{X}})\cdot d\mathbf{\dot{X}}
\label{wz} \\
\Omega _{L}[\mathbf{X}] &=&m\zeta d\mathbf{\dot{X}}\dot{\wedge}d\mathbf{X}-{%
\frac{\zeta ^{2}}{2\mu m}}\mathbf{\dot{X}}\cdot d\mathbf{\dot{X}}\wedge d%
\mathbf{X}-{\frac{\zeta ^{2}}{2\mu m}}\mathbf{X}\cdot d\mathbf{\dot{X}}%
\wedge d\mathbf{\dot{X}} \\
&&-{\frac{\zeta ^{2}}{\mu m}}[\mathbf{X}\cdot d\mathbf{X}\wedge d\mathbf{%
\ddot{X}}+\mathbf{\ddot{X}}\cdot d\mathbf{X}\wedge d\mathbf{X}],  \notag
\end{eqnarray}%
where $\dot{\wedge}$ is as defined in (\ref{wedgedot}). If $\mathbf{X}%
=(X^{1},X^{2},X^{3})$ and $\mathbf{Y}=(Y^{1},Y^{2},Y^{3})$ are two vectors,
we define two exterior products%
\begin{eqnarray*}
d\mathbf{X}\wedge d\mathbf{Y} &=&(dX^{2}\wedge dY^{3}-dY^{2}\wedge
dX^{3},dX^{3}\wedge dY^{1}-dY^{3}\wedge dX^{1}, \\
&&dX^{1}\wedge dY^{2}-dY^{1}\wedge dX^{2}) \\
d\mathbf{X}\wedge d\mathbf{X} &=&2(dX^{2}\wedge dX^{3},dX^{3}\wedge
dX^{1},dX^{1}\wedge dX^{2})\neq 0.
\end{eqnarray*}

The matrix representation of the presymplectic two form $\Omega _{L}$ is
given by%
\begin{equation*}
\Omega _{L}[\mathbf{X}]=\left[
\begin{array}{cccc}
\frac{1}{\mu }\widehat{\mathbf{\ddot{X}}} & -\frac{1}{2}\mathbb{I}-\frac{1}{%
4\mu }\widehat{\mathbf{\dot{X}}} & \frac{1}{2\mu }\widehat{\mathbf{X}} &
\mathbf{0} \\
\frac{1}{2}\mathbb{I}+\frac{1}{4\mu }\widehat{\mathbf{\dot{X}}} & \frac{1}{%
2\mu }\widehat{\mathbf{X}} & \mathbf{0} & \mathbf{0} \\
-\frac{1}{2\mu }\widehat{\mathbf{X}} & \mathbf{0} & \mathbf{0} & \mathbf{0}
\\
\mathbf{0} & \mathbf{0} & \mathbf{0} & \mathbf{0}%
\end{array}%
\right] .
\end{equation*}%
Here, $\mathbb{I}$ denotes $3\times 3$\ identity matrix and we employed the
hat map $\widehat{\mathbf{X}}$ notation
\begin{equation}
\widehat{}:%
%TCIMACRO{\U{211d} }%
%BeginExpansion
\mathbb{R}
%EndExpansion
^{3}\rightarrow \mathfrak{so}\left( 3\right) :\mathbf{X=(}X,Y,Z\mathbf{)}%
\rightarrow \widehat{\mathbf{X}}=\left[
\begin{array}{ccc}
0 & -Z & Y \\
Z & 0 & -X \\
-Y & X & 0%
\end{array}%
\right]  \label{hatmap}
\end{equation}%
which is an isomorphism from $%
%TCIMACRO{\U{211d} }%
%BeginExpansion
\mathbb{R}
%EndExpansion
^{3}$ to the space of skew-symmetric matrices. Note that, the hat map
isomorphism can also be seen by the identity $\widehat{\mathbf{X}}\mathbf{Y}=%
\mathbf{X}\times \mathbf{Y}.$

The Cl\`{e}ment Lagrangian (\ref{LC}) is invariant under translations in $t$
and pseudo-rotations in space. Time translation symmetry gives the
conservation of energy in Eq.(\ref{lcon}). Rotational invariance implies the
conservation of angular momentum
\begin{equation}
\mathbf{J}[\mathbf{X}]=m\mathbf{X}\times \mathbf{\dot{X}}+\frac{m}{2\mu }[2%
\mathbf{X}\times (\mathbf{X}\times \mathbf{\ddot{X}})-\mathbf{\dot{X}}\times
(\mathbf{X}\times \mathbf{\dot{X}})].  \label{momLag}
\end{equation}

\subsection{Dirac-Bergmann Algorithm}

We recall the Darboux' coordinates $(\mathbf{X,\dot{X},P}^{0},\mathbf{P}%
^{1}) $ and compute the Jacobi-Ostrogradsky momenta
\begin{equation}
\mathbf{P}^{0}[\mathbf{X}]=-m\zeta \dot{\mathbf{X}}+\frac{\zeta ^{2}}{\mu m}%
\ddot{\mathbf{X}}\times \mathbf{X},\;\;\;\mathbf{P}^{1}[\mathbf{X}]=\frac{%
\zeta ^{2}}{2\mu m}\mathbf{X}\times \dot{\mathbf{X}},  \label{Leg-Clement}
\end{equation}%
on momentum phase space $T^{\ast }TM$ with the Darboux' coordinates $(%
\mathbf{X,\dot{X},P}^{0},\mathbf{P}^{1})$. The Jacobi-Ostrogradsky momenta $%
\left( \mathbf{P}^{0},\mathbf{P}^{1}\right) $\ presented in (\ref%
{Leg-Clement}) define the Legendre map $\mathcal{F}L:T^{3}M\longrightarrow
T^{\ast }TM$ whose tangent
\begin{equation*}
D\mathcal{F}L=\left[
\begin{array}{cccc}
\mathbb{I} & 0 & 0 & 0 \\
0 & \mathbb{I} & 0 & 0 \\
\frac{1}{\mu }\widehat{\mathbf{\ddot{X}}} & -\mathbb{I} & -\frac{1}{\mu }%
\widehat{\mathbf{X}} & 0 \\
-\frac{1}{2\mu }\widehat{\mathbf{\dot{X}}} & \frac{1}{2\mu }\widehat{\mathbf{%
X}} & 0 & 0%
\end{array}%
\right]
\end{equation*}%
is a $12\times 12$ matrix whose rank is $9$. Note that the momenta $\mathbf{J%
}$ in (\ref{momLag}) can be regarded as the pull-back of the angular
momentum
\begin{equation*}
\mathbf{J=X}\times \mathbf{P}^{0}[\mathbf{X}]+\mathbf{\dot{X}}\times \mathbf{%
P}^{1}[\mathbf{X}].
\end{equation*}

The canonical Hamiltonian function for the Cl\`{e}ment Lagrangian (\ref{LC})
is
\begin{equation}
H^{C}=\mathbf{P}^{0}\cdot \dot{\mathbf{X}}+\mathbf{P}^{1}\cdot \ddot{\mathbf{%
X}}-L^{C}=\frac{m}{2}\zeta \dot{\mathbf{X}}^{2}+\frac{2m\Lambda }{\zeta }+%
\dot{\mathbf{X}}\cdot \mathbf{P}^{0}.  \label{Can-Ham-Clement}
\end{equation}%
The pull-back of the canonical Hamiltonian $H^{C}$ to $T^{2}M$ by
Jacobi-Ostrogradsky momenta corresponds to the energy function $E^{C}[%
\mathbf{X}]$ presented in (\ref{lcon}). We shall apply Dirac constraint
analysis to obtain the Hamiltonian formulation of the dynamics. The
Ostrogradsky momenta $\mathbf{P}^{1}$\ lead to the set $\mathbf{\Phi }$ of
primary constraints
\begin{equation}
\mathbf{\Phi }=\mathbf{P}^{1}-\frac{\zeta ^{2}}{2\mu m}\mathbf{X}\times
\mathbf{\dot{X}.}  \label{Clem-Const-1}
\end{equation}%
The consistency conditions
\begin{equation}
\mathbf{\dot{\Phi}}=\{\mathbf{\Phi },H^{C}+\mathbf{U}\cdot \mathbf{\Phi }%
\}=-m\zeta \mathbf{\dot{X}}-\mathbf{P}^{0}+\frac{\zeta ^{2}}{\mu m}\mathbf{%
U\times X}\approx 0,  \label{Clement-Consistency}
\end{equation}%
of the primary constraints for $\mathbf{\Phi }$ using the Hamiltonian $H^{C}+%
\mathbf{U}\cdot \mathbf{\Phi }$ result the followings. Due to the degeneracy
of the cross product, only two components of the Lagrange multipliers $%
\mathbf{U}$ can be solved from these equations and a secondary constraint
\begin{equation}
\Phi =m\zeta \mathbf{X}\cdot \dot{\mathbf{X}}+\mathbf{X}\cdot \mathbf{P}^{0}
\label{Clem-Const-2}
\end{equation}%
arises. This secondary constraint $\Phi $ can also be derived directly by
the dot product of momenta $\mathbf{P}^{0}$ with $\mathbf{X}$. We add the
secondary constraint $\Phi $ to the Hamiltonian function and define the
total Hamiltonian as
\begin{equation}
H_{T}^{C}=H^{C}+\mathbf{U}\cdot \mathbf{\Phi }+U\Phi .
\label{Clem-total-Ham-imp}
\end{equation}%
Using the total Hamiltonian $H_{T}^{C}$, the consistency conditions give the
set of four equations
\begin{eqnarray}
\dot{\Phi} &=&\{\Phi ,H_{T}^{C}\}=\dot{\mathbf{X}}\cdot \mathbf{B}+\mathbf{U}%
\cdot \mathbf{A}\approx 0,  \label{cc1} \\
\mathbf{\dot{\Phi}} &=&\{\mathbf{\Phi },H_{T}^{C}\}=-m\zeta \dot{\mathbf{X}}-%
\mathbf{P}^{0}+\frac{\zeta ^{2}}{m\mu }\mathbf{U}\times \mathbf{X}-U\mathbf{A%
}\approx 0.  \label{cc2}
\end{eqnarray}%
Here, we used the abbreviations
\begin{equation}
\mathbf{A}=m\zeta \mathbf{X}+\mathbf{P}^{1}\text{ \ \ and \ \ }\mathbf{B}%
=m\zeta \dot{\mathbf{X}}+\mathbf{P}^{0}.  \label{abb}
\end{equation}

Under the assumption of $X^{2}\neq 0$, we solve the Lagrange multipliers $U$
and $\mathbf{U}$ as follows
\begin{eqnarray}
U &=&\frac{-1}{m\zeta X^{2}}\mathbf{X}\cdot \mathbf{B}=\frac{-1}{m\zeta X^{2}%
}\Phi ,  \notag \\
\mathbf{U} &=&\frac{-3}{2m\zeta X^{2}}\mathbf{X}\left( \mathbf{B}\cdot \dot{%
\mathbf{X}}\right) -\frac{\mu m}{\zeta ^{2}X^{2}}\left( \mathbf{B}\times {%
\mathbf{X}}\right)  \label{Clem-Lag-mult}
\end{eqnarray}%
where we used $\mathbf{X}\cdot \mathbf{P}^{1}=0$ from definition of momenta.
Interesting to note that, $U$ vanishes on the constraint submanifold, that
is $U\approx 0$. Note that, since we start with Minkowskian metric, the
condition $X^{2}=0$ refers that the particle is on the light cone. A naive
way for relaxing the condition is to take the limit $X^{2}\rightarrow 0$ in
the expressions. Under this limit, the vector $\mathbf{U}$ approximate the
acceleration $\ddot{\mathbf{X}}$ after the substitution of the momenta. This
observation encourages us to comment on the continuous dependence of the
Lagrange multipliers on the $X^{2}$.

After substitution of the constraints $\Phi $ and $\mathbf{\Phi }$ in (\ref%
{Clem-Const-1} and \ref{Clem-Const-2}), and the Lagrange multipliers $U$ and
$\mathbf{U}$ in (\ref{Clem-Lag-mult}) into implicit form of $H_{T}^{C}$ in (%
\ref{Clem-total-Ham-imp}), we write the total Hamiltonian as%
\begin{eqnarray}
H_{T}^{C} &=&\frac{1}{2}\dot{\mathbf{X}}\cdot \mathbf{P}^{0}-\frac{3}{%
2m\zeta {\mathbf{X}}^{2}}\left( \mathbf{X\cdot P}^{1}\right) \left( \mathbf{B%
}\cdot \dot{\mathbf{X}}\right) -\frac{\mu m}{\zeta ^{2}{\mathbf{X}}^{2}}%
\mathbf{P}^{1}\cdot \left( \mathbf{B}\times {\mathbf{X}}\right)  \notag \\
&&+\frac{1}{2{\mathbf{X}}^{2}}(\mathbf{B\cdot X})(\mathbf{X\cdot \dot{X}})-%
\frac{1}{m\zeta {\mathbf{X}}^{2}}\left( \mathbf{B}\cdot {\mathbf{X}}\right)
^{2}  \label{TH}
\end{eqnarray}%
Using the total Hamiltonian $H_{T}^{C}$, we write the Hamilton's equations
as
\begin{align}
\mathbf{\dot{X}}& \approx \frac{1}{2}\mathbf{\dot{X}}+\frac{\mu m}{\zeta ^{2}%
\mathbf{X}^{2}}\mathbf{P}^{1}\times \mathbf{X}+\frac{1}{2\mathbf{X}^{2}}%
\mathbf{X}(\mathbf{X\cdot \dot{X}})  \notag \\
\mathbf{\ddot{X}}& \approx \frac{\mu m}{\zeta ^{2}\mathbf{X}^{2}}\mathbf{%
X\times B}-\frac{3}{2m\zeta \mathbf{X}^{2}}(\mathbf{B\cdot \dot{X}})\mathbf{X%
}  \notag \\
\mathbf{\dot{P}}^{0}& \approx \frac{\mu m}{\zeta ^{2}\mathbf{X}^{2}}\mathbf{P%
}^{1}\times \mathbf{B}+\frac{3}{2m\zeta \mathbf{X}^{2}}(\mathbf{B\cdot \dot{X%
}})\mathbf{P}^{1}-\frac{1}{2\mathbf{X}^{2}}(\mathbf{\dot{X}\cdot X})\mathbf{B%
}  \notag \\
& -\frac{2}{\mathbf{X}^{4}}\frac{\mu m}{\zeta ^{2}}(\mathbf{P}^{1}\cdot
\mathbf{B\times X})\mathbf{X}  \notag \\
\mathbf{\dot{P}}^{1}& \approx -\frac{1}{2}\mathbf{P}^{0}-\frac{\mu m^{2}}{%
\zeta \mathbf{X}^{2}}\mathbf{P}^{1}\times \mathbf{X}-\frac{m\zeta }{2\mathbf{%
X}^{2}}\mathbf{X}^{i}(\mathbf{X\cdot \dot{X}}).  \label{HamEqCle}
\end{align}%
Here, the first and the last of equations are identically satisfied. The Eq.(%
\ref{HamEq}) gives the weak equality $\mathbf{\ddot{X}}\approx \mathbf{U}$
and the third equation (\ref{HamEqP0}) gives
\begin{equation}
\mathbf{\dot{P}}^{0}=\frac{\zeta ^{2}}{2m\mu }\dot{\mathbf{X}}\times \ddot{%
\mathbf{X}}  \label{HamPo}
\end{equation}%
which, using the definition of $\mathbf{P}^{0}$, results in the
Euler-Lagrange equations (\ref{clee}) for Cl\'{e}ment Lagrangian.

\subsection{Dirac-Poisson Bracket}

All of the four constraints $\mathbf{\Phi }$ and $\Phi $ are second class.
This enables us to define the following nondegenerate $4\times 4$ matrix
\begin{equation}
\mathcal{M}=\left(
\begin{array}{cc}
\{\mathbf{\Phi },\mathbf{\Phi }\} & \{\mathbf{\Phi },\Phi \} \\
\{\Phi ,\mathbf{\Phi }\} & \{\Phi ,\Phi \}%
\end{array}%
\right) =\left(
\begin{array}{cc}
-\frac{\zeta ^{2}}{\mu m}\widehat{\mathbf{X}} & -m\zeta \mathbf{X}-\frac{%
\zeta ^{2}}{2\mu m}\mathbf{X}\times \mathbf{\dot{X}} \\
m\zeta \mathbf{X}^{T}+\frac{\zeta ^{2}}{\mu m}(\mathbf{X}\times \mathbf{\dot{%
X}})^{T} & 0%
\end{array}%
\right)
\end{equation}%
where super script $T$ denoted the transpose of the vectors and the notation
$\widehat{}$ stands for the hatmap in (\ref{hatmap}). The determinant of the
matrix $\mathcal{M}$ is $\zeta ^{6}X^{2}/\mu ^{2}$ and its inverse is
\begin{equation*}
\mathcal{M}^{-1}=\frac{1}{m\zeta X^{2}}\left(
\begin{array}{cc}
\frac{1}{\zeta ^{2}}m\mu (\widehat{\mathbf{X}}+\widehat{\mathbf{P}}^{1}) &
\mathbf{X} \\
-\mathbf{X}^{T} & 0%
\end{array}%
\right) .
\end{equation*}%
Recalling the definition (\ref{diracbrac}) of the Dirac bracket, we construct%
\begin{eqnarray}
\{X^{i},\dot{X}^{j}\}_{DB} &=&\frac{-1}{m\zeta X^{2}}X^{i}X^{j},\text{ \ \ }
\notag \\
\{\dot{X}^{i},\dot{X}^{j}\}_{DB} &=&\frac{-\mu \epsilon ^{ijk}}{\zeta
^{3}X^{2}}A_{k}\text{, \ \ }  \notag \\
\{X^{i},P_{j}^{0}\}_{DB} &=&\delta _{j}^{i}-\frac{\zeta }{2\mu m^{2}X^{2}}%
X^{i}\epsilon _{jkl}\dot{X}^{k}X^{l},  \notag \\
\{\dot{X}^{i},P_{j}^{1}\}_{DB} &=&\delta _{j}^{i}-\frac{1}{2m\zeta X^{2}}%
\left( A_{k}X^{k}\delta _{j}^{i}-A_{j}X^{i}\right) -\frac{1}{\mathbf{X}^{2}}%
X^{i}\delta _{jl}X^{l},  \notag \\
\{\dot{X}^{i},P_{j}^{0}\}_{DB} &=&\frac{1}{2m\zeta X^{2}}\left( \delta
_{j}^{i}\dot{X}^{k}A_{k}-\dot{X}^{i}A_{j}-2X^{i}B_{j}\right) ,  \notag \\
\{P_{i}^{0},P_{j}^{0}\}_{DB} &=&\frac{\zeta }{4\mu m^{2}X^{2}}\left(
\epsilon _{ilj}\dot{X}^{l}\dot{X}^{r}A_{r}-2\epsilon _{isn}\dot{X}%
^{s}X^{n}B_{j}+2\epsilon _{jnr}\dot{X}^{n}X^{r}B_{i}\right) ,  \notag \\
\{P_{i}^{0},P_{j}^{1}\}_{DB} &=&\frac{\zeta }{4\mu m^{2}X^{2}}\left(
\epsilon _{ikl}\dot{X}^{k}X^{l}A_{j}-\epsilon _{jik}\dot{X}%
^{k}A_{r}X^{r}-2m\zeta \epsilon _{rik}\dot{X}^{k}X^{r}\delta
_{jl}X^{l}\right) ,  \notag \\
\{P_{i}^{1},P_{j}^{1}\}_{DB} &=&\frac{-\zeta }{4\mu m^{2}X^{2}}\epsilon
_{jki}X^{k}A_{r}X^{r}  \label{DB-C}
\end{eqnarray}%
where the rest is zero and the abbreviations defined in Eq.(\ref{abb}) are
used. In terms of the Dirac brackets, the equations of motions are
\begin{align}
\mathbf{\dot{X}}=\{{\mathbf{X}},H^{C}\}_{DB}& =\frac{-1}{m\zeta \mathbf{X}%
^{2}}\mathbf{X}(\mathbf{X\cdot B})+\mathbf{\dot{X}} \\
\mathbf{\ddot{X}}=\{\dot{\mathbf{X}},H^{C}\}_{DB}& =\frac{-\mu }{\zeta ^{3}%
\mathbf{X}^{2}}\mathbf{B\times A}-\frac{1}{m\zeta \mathbf{X}^{2}}\mathbf{X}(%
\mathbf{B\cdot \dot{X}}) \\
\mathbf{\dot{P}}^{0}=\{\mathbf{P}^{0},H^{C}\}_{DB}& =\frac{-1}{2m\zeta
\mathbf{X}^{2}}\mathbf{B}(\mathbf{A\cdot \dot{X}})+\frac{1}{2m\zeta \mathbf{X%
}^{2}}\mathbf{A}(\mathbf{B\cdot \dot{X}})  \notag \\
& +\frac{1}{m\zeta \mathbf{X}^{2}}\mathbf{B}(\mathbf{X\cdot B})-\frac{\zeta
}{2\mu m^{2}\mathbf{X}^{2}}\mathbf{\dot{X}\times X}(\mathbf{B_{\cdot }\dot{X}%
})  \label{DBClemHam} \\
\mathbf{\dot{P}}^{1}=\{\mathbf{P}^{1},H^{C}\}_{DB}& =-\mathbf{B}+\frac{1}{%
2m\zeta \mathbf{X}^{2}}\mathbf{B}(\mathbf{A\cdot X})-\frac{1}{2m\zeta
\mathbf{X}^{2}}\mathbf{A}(\mathbf{B\cdot X})+\frac{1}{\mathbf{X}^{2}}\mathbf{%
X}(\mathbf{X\cdot B})
\end{align}%
generated by the canonical Hamiltonian $H^{C}$ in Eq.(\ref{Can-Ham-Clement}%
). Where we used the abbreviations $\mathbf{A}=m\zeta \mathbf{X}+\mathbf{P}%
^{1},\mathbf{B}=m\zeta \mathbf{\dot{X}}+\mathbf{P}^{0}$ and $\mathbf{X}^{2}=%
\mathbf{{X}\cdot X}$. The first and fourth equations are identically
satisfied. The third equation reduces to the Euler-Lagrange equations in the
form of Eq.(\ref{HamPo}). The second equation gives%
\begin{equation}
{\mathbf{X\cdot \ddot{X}+}}\frac{3}{2\mu m^{2}X^{2}}\dot{\mathbf{X}}\cdot
\left( {\mathbf{\ddot{X}\times X}}\right) =\mathbf{0}  \label{CON}
\end{equation}%
which is nothing but the dot product of the Euler-Lagrange Eq.(\ref{clee})
and ${\mathbf{X}}$, hence equal to zero modulo Euler-Lagrange equations.

\subsection{Skinner-Rusk Unified Formalism}

In order to put Cl\'{e}ment dynamics in the form of Skinner-Rusk formalism,
we recall Pontryagin bundle $P^{3}M=T^{3}M\times _{TM}T^{\ast }TM$ and the
presymplectic two-form $\Omega _{P^{3}M}$ in Eq.(\ref{OhmP3M}). The
Hamiltonian function on the presymplectic manifold $\left( P^{3}M,\Omega
_{P^{3}M}\right) $ is%
\begin{equation*}
H_{P^{3}M}=\mathbf{P}^{0}\cdot \mathbf{\dot{X}}+\mathbf{P}^{1}\cdot \mathbf{%
\ddot{X}}+\frac{m\zeta }{2}\dot{X}^{2}+\frac{2m\Lambda }{\zeta }-\frac{\zeta
^{2}}{2\mu m}\mathbf{X}\cdot (\mathbf{\dot{X}}\times \mathbf{\ddot{X}})
\end{equation*}%
and it generates the dynamics according to the Hamilton's equations $%
i_{X_{P^{3}M}}\Omega _{P^{3}M}=-dH_{P^{3}M}$. We recall the general form%
\begin{eqnarray}
X_{P^{3}M} &=&\mathbf{\dot{X}}\cdot \nabla _{\mathbf{X}}+\mathbf{\ddot{X}}%
\cdot \nabla _{\mathbf{\dot{X}}}+\mathbf{\dddot{X}}\cdot \nabla _{\mathbf{%
\ddot{X}}}+\mathbf{C}\cdot \nabla _{\mathbf{\dddot{X}}}  \notag \\
&&+\frac{\zeta ^{2}}{2\mu m}\left( \mathbf{\dot{X}}\times \mathbf{\ddot{X}}%
\right) \cdot \nabla _{\mathbf{P}^{0}}+\left( -m\zeta \mathbf{\dot{X}-}\frac{%
\zeta ^{2}}{2\mu m}\mathbf{X\times \ddot{X}}-\mathbf{P}^{0}\right) \cdot
\nabla _{\mathbf{P}^{1}}  \label{SR-X}
\end{eqnarray}%
with unknown coefficient functions $\mathbf{C}=\left(
C_{1},C_{2},C_{3}\right) $ which will be determined through the algorithm
runs. The graph of the Legendre map $\mathcal{F}L$ is defined by the
constraints
\begin{equation}
\mathbf{\psi }=\mathbf{P}^{0}+m\zeta \mathbf{\dot{X}+}\frac{\zeta ^{2}}{\mu m%
}\mathbf{X\times \ddot{X}}\text{, \ \ }\mathbf{\Phi }=\mathbf{P}^{1}-\frac{%
\zeta ^{2}}{2\mu m}\mathbf{X\times \dot{X}.}
\end{equation}%
The tangency conditions for the constraint functions are
\begin{eqnarray}
X_{P^{3}M}\mathbf{\psi } &=&\mathbf{\psi }_{1}=\frac{3\zeta ^{2}}{2\mu m}%
\mathbf{\dot{X}}\times \mathbf{\ddot{X}+m\zeta \ddot{X}+}\frac{\zeta ^{2}}{%
\mu m}\mathbf{X\times \dddot{X}}  \notag \\
X_{P^{3}M}\mathbf{\Phi } &=&-\mathbf{\psi .}
\end{eqnarray}%
$\mathbf{\psi }_{1}\approx 0$ defines the constraint manifold $W_{1}$. We
need to check tangency condition $X_{P^{3}M}\mathbf{\psi }_{1}\approx 0$ to
decide that whether $W_{0}$ is the final constraint submanifold or not.
Accordingly, we compute
\begin{equation}
X_{P^{3}M}\mathbf{\psi }_{1}=\frac{5\zeta ^{2}}{2\mu m}\mathbf{\dot{X}}%
\times \mathbf{\dddot{X}}+m\zeta \dddot{\mathbf{X}}+\frac{\zeta ^{2}}{\mu m}%
\mathbf{X\times C.}  \label{secCl}
\end{equation}%
By requiring $X_{P^{3}M}\mathbf{\psi }_{1}$ be zero, we obtain following
information. First of all, the rank of the matrix $\mathbf{\hat{X}}$ is $2$
(except from $\mathbf{X=0}$) out of $3,$ hence solving $\mathbf{C}$ uniquely
from $X_{P^{3}M}\mathbf{\psi }_{1}=\mathbf{0}$ is not possible. We can only
get two components of $\mathbf{C}=\left( C_{1},C_{2},C_{3}\right) $ in terms
of the third one, say we solve $C_{1}$ and $C_{2}$ in terms of $C_{3}$. In
addition, from Eq.(\ref{secCl}), we get a single secondary constraint
\begin{equation*}
\psi _{2}=m\zeta \mathbf{X}\cdot {\dddot{\mathbf{X}}}+\frac{5\zeta ^{2}}{%
2\mu m}\mathbf{X\cdot \dot{X}\times \dddot{X}}
\end{equation*}%
which is obtained by taking dot product of Eq.(\ref{secCl}) with $\mathbf{X}$%
. Hence, we arrive at the submanifold $W_{2}$ which is the final constraint
submanifold since the requirement that
\begin{equation}
X_{P^{3}M}\psi _{2}=\mathbf{\dddot{X}\cdot }\left( m\zeta \mathbf{\dot{X}+}%
\frac{5\zeta ^{2}}{2\mu m}\mathbf{X\times \ddot{X}}\right) +\mathbf{C\cdot }%
\left( m\zeta \mathbf{X+}\frac{5\zeta ^{2}}{2\mu m}\mathbf{X\times \dot{X}}%
\right) =0  \label{B3}
\end{equation}%
leads us to determine $C_{3}$ and substitute in $C_{1}$ and $C_{2}$. Thus,
\begin{align}
\mathbf{C}& =\frac{-1}{m\zeta X^{2}}\mathbf{X}(\mathbf{\dddot{X}\cdot }%
(m\zeta \mathbf{\dot{X}+}\frac{5\zeta ^{2}}{2\mu m}\mathbf{X\times \dot{X}}))
\notag \\
& +\frac{\mu }{\zeta ^{3}X^{2}}(m\zeta \mathbf{X+}\frac{5\zeta ^{2}}{2\mu m}%
\mathbf{X\times \dot{X}})\times (m\zeta \mathbf{\dddot{X}+}+\frac{5\zeta ^{2}%
}{2\mu m}\mathbf{\dddot{X}\times \dot{X}}).  \label{C}
\end{align}

Note that, the substitutions of Legendre map in $H_{P^{3}M}$ leads to the
Lagrangian energy
\begin{equation}
E_{T^{3}M}=-{\frac{m\zeta }{2}}\dot{X}^{2}+{\frac{\zeta ^{2}}{\mu m}}\mathbf{%
X}\cdot (\mathbf{\dot{X}}\times \mathbf{\ddot{X}})+\frac{2m\Lambda }{\zeta }
\label{enCle}
\end{equation}%
on $T^{3}M$. On the projection $S_{f}=pr_{1}\left( W_{f}\right) $ of $W_{f}$
in (\ref{fscl}), the Hamilton's equations are given by means of the
presymplectic relation%
\begin{equation}
\left. i_{X_{T^{3}M}}\Omega _{T^{3}M}+dE_{T^{3}M}\right\vert _{S_{f}}=0,
\label{PreSymp}
\end{equation}%
where $\Omega _{T^{3}M}$ is the presymplectic structure on $T^{3}M$ in Eq.(%
\ref{OhmClLag}). Here,
\begin{equation}
X_{T^{3}M}=\left( pr_{1}\right) _{\ast }X_{P^{3}M}=\mathbf{\dot{X}}\cdot
\nabla _{\mathbf{X}}+\mathbf{\ddot{X}}\cdot \nabla _{\mathbf{\dot{X}}}+%
\mathbf{\dddot{X}}\cdot \nabla _{\mathbf{\ddot{X}}}+\mathbf{C}\cdot \nabla _{%
\mathbf{\dddot{X}}}.  \label{SR-C}
\end{equation}%
From presymplectic relation (\ref{PreSymp}), we can derive the
Euler-Lagrange equations (\ref{clee}).

\section{Conclusions}

In the present work, we have derived the Hamiltonian and the Skinner-Rusk
unified formalisms for the second order degenerate Sar\i o\u{g}lu-Tekin and
Cl\`{e}ment Lagrangians. We have applied the Dirac-Bergmann constraint
algorithm in order to arrive at the Hamiltonian pictures on the momemtum
phase spaces whereas we have applied the Gotay-Nester-Hinds algorithm while
investigating the Skinner-Rusk unified formalisms on the proper Whitney
bundles.\ As a result, for the Sar\i o\u{g}lu-Tekin Lagrangian (\ref{stlag}%
), we have obtained the total Hamiltonian function in (\ref{TH-SK}), the
Hamilton's equations in Eq.(\ref{HamEqST1}-\ref{HamEqST2}), the
Dirac-Poisson bracket in Eqs.(\ref{DB-SK}), and the vector field generating
the unified formalism in Eq.(\ref{X-SK}). For the Cl\`{e}ment Lagrangian, we
have calculated the constraint Hamiltonian function in Eq.(\ref{TH}), the
Hamilton's equations in Eqs.(\ref{HamEqCle}), the Dirac-Poisson bracket in
Eqs.(\ref{DB-C}), and the vector field generating the unified formalism in
Eq.(\ref{SR-C}).

Here is the list of some possible complementary and future works:

\begin{itemize}
\item We are planning to study the Hamiltonian formalism of Sar\i o\u{g}%
lu-Tekin and Cl\`{e}ment Lagrangian after writing them as first order
Lagrangians by properly defining new coordinates and Lagrange multipliers.
By this, we can able to make a comparative study of the Hamiltonian
representations of the degenerate second order Lagrangians and their reduced
degenerate first order equivalents on some concrete problems.

\item Schmidt-Legendre transformation is an alternative method while
reducing a second order Lagrangian function to the first order. It works
both for non-degenerate and degenerate systems \cite{Sc94,Sc95}. In this
theory, the acceleration is assumed as a new coordinate instead of the
velocity \cite{AnGoMaMa10,AnGoMa07,EsGu16}. We are planning to make a detail
analysis of the present discussions in terms of the Schmidt-Legendre
transformation.

\item Both of the Sar\i o\u{g}lu-Tekin and Cl\`{e}ment Lagrangians have
rotational symmetry. In \cite{GaHoRa11}, the higher dimensional version of
the Lagrangian reduction \cite{CeMaRa01,MaRa98} has been presented. By
motivating this, we are planning to exhibit formal reductions of the Sar\i o%
\u{g}lu-Tekin and Cl\`{e}ment Lagrangians under rotational symmetry.
\end{itemize}

\bigskip

\bigskip

\bigskip

\bigskip

\end{document}